\documentclass[10pt,conference,review,anonymous]{IEEEtran}
\IEEEoverridecommandlockouts

\usepackage{cite}
\usepackage{amsmath,amssymb,amsfonts}
\usepackage{graphicx}
\usepackage{textcomp}
\usepackage{xcolor}
\usepackage{algpseudocode}
\usepackage{algorithm}
\usepackage{xspace}
\usepackage{listings}
\usepackage{amssymb}
\usepackage{color}
\usepackage{multirow}
\usepackage{dblfloatfix}
\usepackage{lscape} 
\usepackage{xparse}
\usepackage{placeins}
\usepackage{caption} 
\usepackage{subcaption}
\usepackage{enumitem}
\usepackage{array, makecell}
\usepackage{wrapfig}
\usepackage{colortbl}
\usepackage{tabularx} 
\usepackage{booktabs}
\usepackage{courier}
\usepackage{ragged2e}
\usepackage{amsthm}
\usepackage{hyperref}
\usepackage{comment}
\usepackage{pdflscape}
\usepackage{afterpage}
\usepackage{fontawesome}
\usepackage{pifont}
\usepackage{lscape}
\usepackage{enumitem}

\setlist[itemize]{noitemsep, topsep=4pt}
\definecolor{lightgray}{HTML}{f6f6f6}
\definecolor{darkgray}{rgb}{.4,.4,.4}
\definecolor{darkblue}{HTML}{800080}
\definecolor{brickred}{HTML}{b04f4f}
\definecolor{purple}{rgb}{0.65, 0.12, 0.82}
\definecolor{diffadd}{HTML}{288f26}
\definecolor{diffrmbg}{HTML}{ffebe9}
\definecolor{diffaddbg}{HTML}{e6ffeb}
\definecolor{diffremove}{HTML}{de4f54}
\definecolor{carrotorange}{rgb}{0.8, 0.33, 0.0}
\definecolor{highlight}{HTML}{fefbc2}
\definecolor{bluegray}{HTML}{3182bd}

\lstdefinelanguage{JavaScript}{
  keywords={typeof, new, true, false, catch, function, return, null, catch, switch, var, const, let, extends, if, in, while, do, else, case, break, async, await, of},
  keywordstyle=\color{darkblue}\bfseries,
  ndkeywords={class, export, boolean, throw, implements, import, this, setTimeout},
  ndkeywordstyle=\color{brickred}\bfseries,
  identifierstyle=\color{black},
  sensitive=false,
  comment=[l]{//},
  morecomment=[f][\color{diffadd}\bfseries]{+\ },
  morecomment=[s]{/*}{*/},
  morecomment=[f][\color{diffremove}\bfseries]{- },
  commentstyle=\color{violet}\ttfamily,
  stringstyle=\color{carrotorange}\ttfamily,
  morestring=[b]',
  morestring=[b]"
}

\lstdefinelanguage{Python}{
  keywords={typeof, new, true, false, catch, function, return, null, catch, switch, var, const, let, extends, if, in, while, do, else, case, break, async, await, of, from, import, class, def},
  keywordstyle=\color{darkblue}\bfseries,
  ndkeywords={class, export, boolean, throw, implements, import, this, setTimeout, self, __init__},
  ndkeywordstyle=\color{brickred}\bfseries,
  identifierstyle=\color{black},
  sensitive=false,
  comment=[l]{//},
  morecomment=[f][\color{diffadd}\bfseries]{+\ },
  morecomment=[s]{/*}{*/},
  morecomment=[f][\color{diffremove}\bfseries]{- },
  commentstyle=\color{violet}\ttfamily,
  stringstyle=\color{carrotorange}\ttfamily,
  morestring=[b]',
  morestring=[b]"
}

\theoremstyle{definition}

\newcommand{\header}[1]{\par\smallskip\noindent\textbf{#1.}}

\def\BibTeX{{\rm B\kern-.05em{\sc i\kern-.025em b}\kern-.08em
    T\kern-.1667em\lower.7ex\hbox{E}\kern-.125emX}}
    
\newboolean{showcomments}
\setboolean{showcomments}{true}
\ifthenelse{\boolean{showcomments}}
{
	\definecolor{myyellow}{RGB}{255, 228, 26}
	\definecolor{myblue}{RGB}{50, 50, 220}
	\newcommand{\nb}[2]{
		{\sf
			\fcolorbox{myyellow}{yellow}{\scriptsize\textbf{#1}}%
			$\blacktriangleright$%
			{\color{myblue}\fontsize{7pt}{8pt}\selectfont\textbf{#2}}%
		}%
	}
}
{
	\newcommand{\nb}[2]{}
}

\newcommand{\toolname}{\textsc{Lance}\xspace}

\newcommand{\benchmark}{\textsc{APIeval}\xspace}
\newcommand{\benchmarkfullname}{\textsc{API Evaluation}\xspace}

\newcommand{\code}[1]{{\small\ttfamily\texttt{#1}}}

\newcommand{\incoder}{\textsc{InCoder}\xspace}
\newcommand{\codetfive}{\textsc{CodeT5}\xspace}

\newcommand{\copilotfullname}{\textsc{Copilot}\xspace}
\newcommand{\copilot}{\textsc{Copilot}\xspace}
\newcommand{\codeshisperer}{\textsc{CodeWhisperer}\xspace}
\newcommand{\gptthree}{\textsc{GPT-3.5}\xspace}
\newcommand{\gpt}{\textsc{GPT-4}\xspace}
\newcommand{\llama}{\textsc{Llama 2}\xspace}

\makeatletter
\DeclareRobustCommand{\change}{%
  \@bsphack
  \leavevmode
  \color{blue}
  \@esphack
}
\DeclareRobustCommand{\stopchange}{%
  \@bsphack
  \normalcolor
  \@esphack
}
\makeatother

\begin{document}

\title{
Contextual API Completion for Unseen Repositories Using LLMs
}

\makeatletter
\newcommand{\linebreakand}{%
  \end{@IEEEauthorhalign}
  \hfill\mbox{}\par
  \mbox{}\hfill\begin{@IEEEauthorhalign}
}

\makeatother
\author{
    \IEEEauthorblockN{Noor Nashid}
    \IEEEauthorblockA{
        \textit{University of British Columbia}\\
        Vancouver, Canada \\
        nashid@ece.ubc.ca
    }
    \and
    \IEEEauthorblockN{Taha Shabani}
    \IEEEauthorblockA{
        \textit{University of British Columbia}\\
        Vancouver, Canada \\
        taha.shabani@ece.ubc.ca
    }
    \linebreakand
    \IEEEauthorblockN{Parsa Alian}
    \IEEEauthorblockA{
        \textit{University of British Columbia}\\
        Vancouver, Canada \\
        palian@ece.ubc.ca
    }
    \and
    \IEEEauthorblockN{Ali Mesbah}
    \IEEEauthorblockA{
        \textit{University of British Columbia}\\
        Vancouver, Canada \\
        amesbah@ece.ubc.ca
    }
}

\maketitle
\thispagestyle{plain}
\pagestyle{plain}

\begin{abstract}

Large language models have made substantial progress in addressing diverse code-related tasks. However, their adoption is hindered by inconsistencies in generating output due to the lack of real-world, domain-specific information, such as for intra-repository API calls for unseen software projects. We introduce a novel technique to mitigate hallucinations by leveraging global and local contextual information within a code repository for API completion tasks. Our approach is tailored to refine code completion tasks, with a focus on optimizing local API completions. We examine relevant import statements during API completion to derive insights into local APIs, drawing from their method signatures. For API token completion, we analyze the inline variables and correlate them with the appropriate imported modules, thereby allowing our approach to rank the most contextually relevant suggestions from the available local APIs. Further, for conversational API completion, we gather APIs that are most relevant to the developer query with a retrieval-based search across the project. We employ our tool, \toolname, within the framework of our proposed benchmark, \benchmark, encompassing two different programming languages. Our evaluation yields an average accuracy of 82.6\% for API token completion and 76.9\% for conversational API completion tasks. On average, \toolname surpasses \copilot by 143\% and 142\% for API token completion and conversational API completion, respectively. The implications of our findings are substantial for developers, suggesting that our lightweight context analysis can be applied to multilingual environments without language-specific training or fine-tuning, allowing for efficient implementation with minimal examples and effort.
\end{abstract}





\section{Introduction}

Developer tools such as \codeshisperer~\footnote{https://aws.amazon.com/codewhisperer} and \copilotfullname~\footnote{https://copilot.github.com}, built on top of Large Language Models (LLMs),  have gained significant popularity recently. Since LLMs, such as \gpt~\cite{gpt4:openai:arxiv23} and \llama~\cite{touvron:llama2:arxiv23}, are trained on massive corpora of code, they are capable of generating syntactically correct code that aligns with the intention of the developer. The effectiveness of LLM-assisted developer tools in aiding a broad spectrum of coding tasks~\cite{denny:conversing-with-copilot:sigcse23, ross:conversational-interaction-for-coding:iui23, xie:chatunitest:arxiv23, tian:chatgpt-programming-assistance:arxiv2023, xia:chatrepair:arxiv2023, lertbanjongam2022empirical, pearce:copilot-security:2021, nhan:copilot-empirical-evaluation:msr22, vaithilingam:copilot:chi22, dakhel2022github, saki:copilot-pair-programming:icse22, cedar} has led to notable improvements in software development productivity~\cite{ziegler:measuring-github-copilot-productivity:acm24}.

Despite their strengths, they still face significant challenges within unseen software repositories. While LLMs can handle popular software libraries, domain-specific API tasks in unseen projects require a deeper understanding of unique functionalities and custom logic defined within the project's code repository. This gap underlines the need to adapt LLMs to project-specific intra-repository contextual information, a capability not present in their initial training dataset. LLMs need to incorporate such domain-specific information for fact-based output generation to avoid hallucinations~\cite{ji:surveyhallucination:22}. 

Techniques such as fine-tuning can be employed to incorporate repository-specific contextual information. However, fine-tuning existing LLMs can be prohibitively expensive. Closed-source language models, such as \gpt, are not readily available for fine-tuning, further limiting the feasibility of this approach. The dynamic nature of software development, with frequent updates to code across files, complicates maintaining an updated model, necessitating ongoing adjustments. As a result, a language model would need to be continually fine-tuned to reflect new project code changes.
%

%
Furthermore, privacy concerns may deter the sharing of sensitive private code for fine-tuning purposes, as the risk of exposing proprietary information or confidential data is high.  Also, granting AI-assisted tools, such as \copilot, direct access to the entire repository would expose the full scope of the project's code to an external service.
In addition, security concerns remain a significant challenge~\cite{breaux:privacy-and-security-requirements:tse07, alhirabi:security-and-privacy-requirements:acm21}, given that proprietary code repositories are company-specific and intended for use solely within that specific organization. These challenges underscore the need for solutions that can enhance LLMs to handle project-specific contexts without the need to share entire code repositories or fine-tune the model.

LLMs have demonstrated the ability to adapt to new tasks without the need for task-specific fine-tuning, employing a method known as prompt-based learning~\cite{prompt-learning-survey:liu:21}. During an API completion task, the prompt consists of \emph{incomplete code} that the developer intends to complete, in addition to any optional contextual information~\cite{xu:polycoder:maps22}. 

Our insight is that project-specific intra-repository context can enhance LLMs in the broader context beyond the initially provided incomplete code;  source code with well-defined semantics enables deterministic analyses to extract relevant information. 
By incorporating a program analysis step, we can extract project-specific cross-file contextual information, allowing the underlying models to directly access all the relevant code and the intended usage without ambiguity. However, given the specificity and diversity of programming languages, there is a need for a language-agnostic program analysis step. This becomes increasingly important in the context of LLMs~\cite{codex-model:22, li:starcoder:arxiv23, openai-models}, as they can handle multiple programming languages. We conjecture that a lightweight, language-agnostic analysis step, designed to extract relevant contextual information --- albeit not as precise as traditional program analysis techniques~\cite{soot} --- could be instrumental for such models. Based on these insights, we propose \toolname (\textbf{L}ightweight \textbf{AN}alysis for \textbf{C}ontext \textbf{E}xtraction), an approach that extracts relevant cross-file contextual information at the intra-repository level through lightweight, language-agnostic static analysis. 


Recognizing the need for comprehensive evaluation of LLMs in software development, current benchmarks~\cite{chen:humaneval:arxiv21, austin:program-synthesis-and-llm:arxiv21, cassano:benchmarking-neural-code:23, cassano:benchmarking-neural-code:23, zheng:humaneval-x:kdd23} despite their valuable contributions, do not fully encapsulate the complexities of real-world software development scenarios, particularly in the context of API usage for projects not previously encountered by the models. To address these gaps, we introduce  \benchmark (API Evaluation), a multilingual benchmark designed for a more nuanced assessment of LLMs, focusing on cross-file API interactions. 

In this paper, we make the following contributions:

\begin{itemize}

\item A technique, called \toolname, designed to improve the accuracy of unseen project-specific API completions using LLMs. It incorporates context analysis to facilitate both next-token API completion and conversational API completion.

\item A lightweight static analysis using language-agnostic syntax trees to ensure the applicability of our approach across programming languages. It extracts information relevant to the API completion through global contextual analysis using import statements, and inline variable analysis for local context. 

\item The development of a new multilingual benchmark specifically tailored for evaluating cross-file API invocation tasks with LLMs. 

\item An empirical evaluation comparing against \copilot code completion and LLMs, namely \gptthree and \gpt, with local file context. On average, for API token completion, \toolname outperforms \copilot by 142.53\% and LLMs with local file only context by 181.54\%. Similarly, for conversational API completion, \toolname outperforms \copilot and LLMs with local file context by a significant margin of 141.82\% and 208.16\%, respectively.


\end{itemize}

This work has implications for practitioners and tool designers. Our results show that through the integration of lightweight program analysis for context selection to construct contextual prompts, \toolname can facilitate API completion tasks without necessitating extensive task-specific data collection, model training, or fine-tuning. 

\section{Motivation} 
\label{sec:motivation}
APIs enable developers to leverage existing libraries and frameworks, which streamlines the development process and enhances code quality~\cite{bloch:effective-java:2008, peng:api-survey:tse2023}. APIs also establish the contract for utilizing various functions or methods, specifying the types of calls that can be made, how to make them, and the appropriate arguments that need to be provided. 
Tools such as \copilot, powered by LLMs, have proven to be effective in handling complex coding tasks, including API code completion, suggesting an evolution towards an intelligent coding environment~\cite{vaithilingam:copilot:chi22, ziegler:copilot-productivity:maps:22, dakhel2022github}. 
However, there have been concerns about the limitations and potential risks of LLMs, as they have been observed to produce output that is nonsensical or unfaithful to the source input~\cite{bender:stochastic-parrots:21, liu:hallucination-detection-benchmark:22}. 
Currently, LLM-assisted code completion is typically prompted based on the previous code tokens from the same file~\cite{xu:polycoder:maps22}. As a result, feeding code tokens to a large language model in this ad-hoc fashion can lead to the generation of plausible-looking output that is a \emph{hallucination} rather than the desired output. This can occur if the model does not have sufficient context to generate accurate and coherent results. Such hallucinations are especially prevalent in tasks such as project-specific API completion requiring a better understanding of project-related contexts. 

As a running example, we will use Listing~\ref{lst:class-with-cross-file-context}, demonstrating a Python class implementation that leverages several dependencies across different files within a private code repository. This example demonstrates a common practice in software development of importing and integrating functionalities from various classes and modules.
\begin{lstlisting}[language=Python, caption={Hotel management system with cross-file context},label={lst:class-with-cross-file-context}]
# hotel_management_system.py
from datetime import datetime, timedelta
import database_helper as dbh
import text_processing as tp
import payment_processing as pp
from hotel import Hotel
from user import User
from review import Review

class HotelReviewAndBookingSystem:
  def __init__(self, hotel_name, rooms):
    self.hotel = Hotel(hotel_name, rooms)    

  def analyze_sentiments(self):
    sentiments = self.hotel.reviews()
    for review in self.reviews:
      review_text = review.text
      sentiment = tp. # API token complete task
-     sentiment = tp.analyze(review_text) # Copilot
+     sentiment = tp.sentiment_analysis(review_text)                           
  ...
  
  def bookroom(self, type, number, name, payment):
    result = self.hotel.book(type, number, guest)

    if result == "Success!":
      # how to process payment with PaymentProcessor? 
-     pp.process_payment(payment) # Copilot
+     pp.process_payment(name, payment) 
\end{lstlisting}
\header{Next Token API Completion} Listing~\ref{lst:class-with-cross-file-context} line 18 presents an example of next token API completion; a developer is implementing a function in a project, where the API code to be completed uses a project-specific API. Completing this API call requires project-level insights, extending beyond the immediate code in the file, to understand dependencies across multiple files some of which are imported. This cross-file context analysis is essential for accurate API token completion. 

If the developer uses \copilot to assist with the API completion, particularly for API calls in an unseen software project, the limitations of current LLM-based developer tools become apparent. The suggestions provided by the LLM, which are based on general coding patterns, lack the specific context needed to complete the code. For such an API completion task, \copilot without a contextual understanding of the specific \code{text\_processing} library employed within the project might suggest a non-existent function named \code{analyze} (line-19). Line 20 shows the correct API completion, namely \code{sentiment\_analysis}. Listing~\ref{lst:text-processing-file-context} shows the functions in \code{text\_processing} available for this API invocation, which includes \code{sentiment\_analysis}. The detection and inclusion of such contextual information can provide the underlying LLM with the required project-specific contextual information.  

\begin{lstlisting}[language=Python, caption={Text processing utility functions},label={lst:text-processing-file-context}]
# text_processing.py
from typing import List, Dict
import datetime

def tokenize(text: str) -> List[str]
def count_words(text: str) -> int
def find_entities(text: str) -> List[str]
def get_word_frequency(text: str, word: str = None) -> Dict[str, int]
def stem(text: str) -> str
def sentiment_analysis(text: str) -> str
def lemmatize(text: str) -> str
def spell_check(text: str) -> str
def translate(text: str, target_language: str) -> str
def get_synonyms(word: str) -> List[str]
...
\end{lstlisting}

\header{Conversational API Completion}   
Leveraging natural language queries during coding tasks is becoming increasingly popular~\cite{yin:general-purpose-code-generation:acl17, le:srevey-dl4se:csur20, li:starcoder:arxiv23, touvron:llama2:arxiv23, gpt4:openai:arxiv23} and the notion of conversational interaction during programming~\cite{ym:conversational-programming:arxiv23, mcnutt:ai-powered-code-assistants:chi23, ross:conversational-interaction-for-coding:iui23} has advanced rapidly in recent years. 

As a result, the use of natural language queries to facilitate API invocations during programming is becoming increasingly relevant. This is shown in line 27 of Listing~\ref{lst:class-with-cross-file-context}. Here, the developer employs a natural language query to leverage \copilot to complete the API code within the context of the project. Despite the query providing clear information about the intent of the developer, which could guide \copilot towards an accurate API completion, the generated code (line 28) is incorrect. 
\begin{lstlisting}[language=Python, caption={PaymentProcessor class.},label={lst:payment_processor}]
# payment_processor.py
import payment import Payment 
...
def process_payment(name: str, payment: Payment) -> bool
def refund_payment(transaction_id: str) -> bool
\end{lstlisting}
Listing~\ref{lst:payment_processor} shows that \code{process\_payment} takes two arguments (not one as generated by \copilot). This inconsistency highlights the difficulties faced by current LLM-based tools in interpreting and executing conversational cues in the absence of detailed project-specific context.

We hypothesize that incorporating a program analysis step could be helpful in systematically gathering project-specific contextual information. This step can provide the underlying LLM with the details necessary to produce more accurate and contextually relevant code suggestions.
\section{Approach}
\label{sec:approach}

\begin{figure*}[h]
	\centering
        \includegraphics[width=\textwidth]{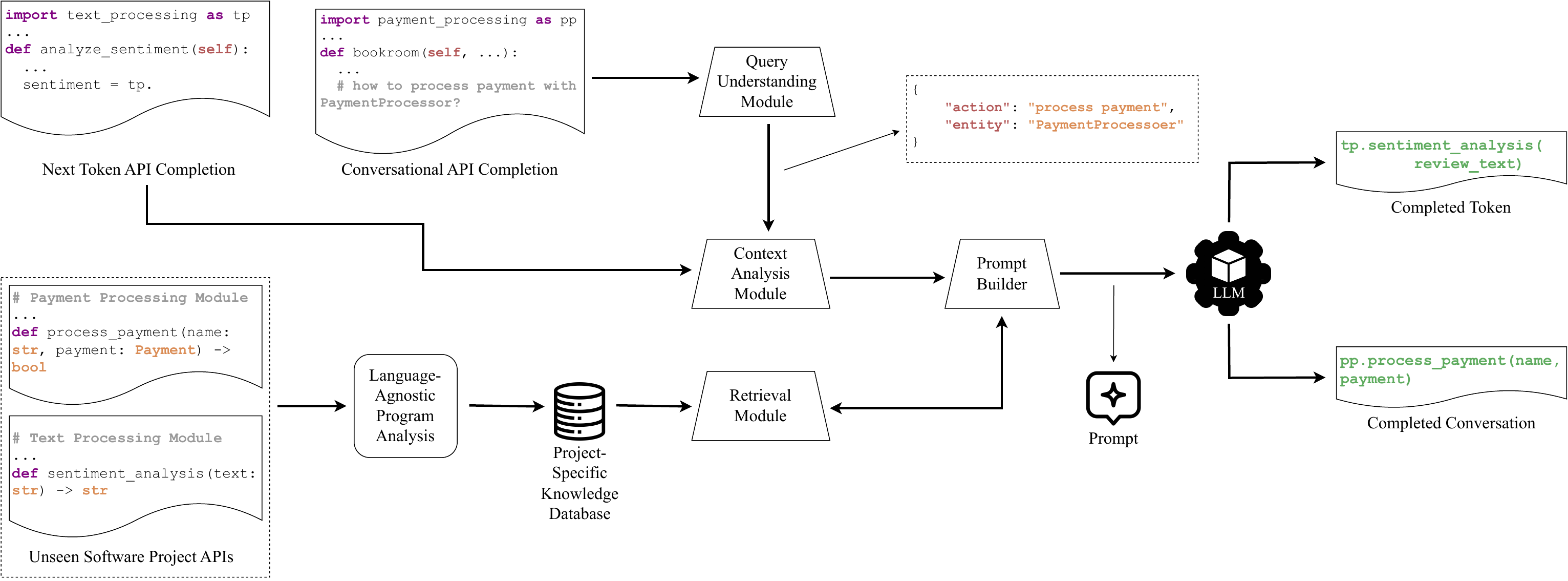}
	\vspace{0em}
	\caption{Overview of our approach \toolname.}
	\vspace{-1.2em}
	\label{fig:overview}
\end{figure*}

Figure \ref{fig:overview} depicts the overall architecture of our approach \toolname. We incorporate a context analysis phase to identify all instances where the code completion task involves a project-specific API. After the extraction of cross-file context, we construct context-augmented prompts to guide the underlying LLM in producing API completion suggestions pertinent to the unseen software project.

In the context analysis phase, our objective is to extract all API-related information across a project. This entails a systematic identification and extraction of function and method calls along with their names, parameters, and code documentation, if available. We also extract class definitions and gather their names, inheritance structures, and descriptive class-level documentation, if available. Given an unseen project, our analysis step aims to gather all the API-related information. An API function is denoted as $f$ and is formally defined as a tuple $(n_f, v_f, P, c_f, r_f)$. The tuple encompasses several elements. The identifier $n_f$ (Name) allows the function to be invoked. The element $v_f$ (Visibility) indicates the function's access scope, such as public or private. In Java, this can be access modifiers like \texttt{public}, \texttt{private}, and \texttt{protected}. In contrast, Python does not implement access modifiers in the same way. The $P$ (Parameters) component consists of a sequence of tuples. Each tuple describes a function's input with a name ($n_p$), type ($t_p$), and an optional comment ($c_p$) if available. The function can accept a variable number of parameters as defined in its declaration. We collect comments for that class ($c_f$), if available. Finally, $r_f$ (Return Type) defines the type of value that the function returns. Java requires an explicit declaration of the return type, including \texttt{void} for no return value. Conversely, Python implicitly returns \texttt{None} when no return statement is encountered. Type information is collected only if it is explicitly present in the source code. 

\subsection{Construction of Project-Specific Context}
To extract API-related information, our approach leverages Tree-sitter~\cite{tree-sitter} to obtain language-agnostic abstract syntax trees (AST). 
We traverse all files within the project to aggregate a detailed dataset of metadata regarding API functions ($f$). 
For every API, we generate embedding for the name of that API ($n_f$) using ADA~\cite{ada} text embedding. 
These embeddings are indexed within a vector database, enabling rapid querying and assessment of semantic similarities between API functions during conversational API queries.

For \code{text\_processing} library in Listing~\ref{lst:text-processing-file-context}, we generate embeddings for the module name and for each defined function, including \code{sentiment\_analysis}. This step ensures that embeddings are generated for each API function identified within the unseen software project. 

\subsection{Context Analysis for Next Token API Completion}
For the next token API completion task, we examine the module name involved in the API call. Following the identification of the module name, we proceed to find all method definitions available for that module. This serves as the basis for all the method signature-related context information available for the next token completion task. We conduct further analysis if the code statement initiates with an assignment statement. In cases where the statement begins with an assignment, the identifier name of the assignment statement can provide further local contextual cues. We rank the available method definitions for the module by assessing the similarity between the identifier name and the method name ($n_f$). This ranking is based on embedding similarity, where method definitions are ranked according to their semantic resemblance to the identifier name, leveraging the embeddings generated in the previous step.

In Listing~\ref{lst:class-with-cross-file-context}, when we encounter the function invocation in line 18, we examine AST to determine the origin of this API call. The analysis of import statements using AST reveals that $tp$ is an aliased import from \code{text\_processing} module. Then, we extract all method signatures from this module to identify available methods for invocation. A further local analysis at the statement level using AST involves identifying variables used up to the point of the current development focus. In the assignment statement, \texttt{sentiment}, the identifier name, serves as an additional local contextual cue. Method signatures are then ranked based on their relevance with the identifier name. This ranking leverages embeddings generated in the previous step, emphasizing semantic similarity between the identifier name and the available method names. These analysis steps are based on TreeSitter~\cite{tree-sitter} and do not rely on language-specific features. 


\subsection{Context Analysis for Conversational API Completion}
For a given conversational API query, the goal is to understand the intention of the developer, such as what code entity the developer is attempting to invoke and to determine what operation they are attempting to perform.

For instance, in the conversational API completion example illustrated in Figure \ref{fig:overview}, the developer seeks assistance from \copilot with query \code{how to process payment with PaymentProcessor?}. To interpret this query, we employ an LLM to parse the natural language request. The objective here is to extract two key pieces of information: the name of the code entity and the specific operation the developer seeks to implement. We use LLM to leverage their advanced natural language processing capabilities to bridge the gap between human conversational requests and the structured requirements of an API completion task. After the identification of the entity (\texttt{PaymentProcessor}) and operation (\texttt{process payment}), the next step is to match these identified elements with the project-specific metadata we have previously indexed. This step serves as the basis for conversational query will be completed in the unseen project context. Next, we perform an embedding similarity analysis between the named entity in the query and the entities indexed within the project. This analysis identifies the project entities most relevant to the query, aiming to match the developer's intent with the existing code entities in the project. We then rank the method signatures associated with the identified code entity based on their relevance to the specified operation. This ranking process relies on evaluating the semantic similarity between the operation described in the developer query and the operations supported by the methods available in the code entity within the project. The goal is to prioritize method signatures that most closely align with the intended action from the developer.

In our example, this approach initially identifies the code entity (\texttt{PaymentProcessor}) and searches for a match in the embedding database. The developer query might refer to this entity in various forms, such as \code{Payment Processor}, or include typos. However, the use of embedding aids in matching similar names despite such variations. This ensures that even with differently phrased queries or minor errors, we can find relevant entities. At the end of this step, we have a ranked list of code entities that are contextually relevant to the developer query.

\subsection{Context-Augmented Prompt Construction}
Following the identification and ranking of relevant code entities and operations, the next phase involves the construction of context-augmented prompts. These prompts are designed to incorporate the contextual information extracted earlier to guide the underlying LLM to relevant API completion suggestions. The prompt structure prioritizes the top-ranked code entities, as determined by our embedding similarity analysis. This ranked inclusion helps the LLM focus on the most contextually relevant entities. The constructed prompt is fed into the LLM, and the output from the LLM is then used for evaluation.

\section{APIEval Benchmark}
\label{sec:apieval}
The goal of developing the \benchmark benchmark is to design tasks that closely mimic real-world software development scenarios. It specifically focuses on API usage and cross-file interactions in contexts unfamiliar to LLMs to mitigate data leakage, thereby challenging these models with tasks they have not encountered.

\header{Rationale} As the proliferation of LLMs continues, the challenge of effectively evaluating their capabilities has become increasingly apparent. Prominent code evaluation benchmarks, namely MBPP~\cite{austin:program-synthesis-and-llm:arxiv21} and HumanEval~\cite{chen:humaneval:arxiv21}, focused on the generation of Python code from natural language descriptions. Subsequently, MultiPL-MBPP~\cite{cassano:benchmarking-neural-code:23} and MultiPL-HumanEval~\cite{cassano:benchmarking-neural-code:23} extended the evaluation to eighteen additional programming languages, aiming to assess the performance of LLMs across a broader spectrum of programming languages. HumanEval-X~\cite{zheng:humaneval-x:kdd23} enhanced the HumanEval benchmark by incorporating multiple test cases. These benchmarks primarily assess the ability of LLMs to generate a function or a statement based on given natural language descriptions. However, they fall short of evaluating the performance of LLM in real-world developer settings, where developers often need to reference and integrate multiple modules for code generation. Another significant challenge in the effective evaluation of LLMs is the issue of potential data leakage~\cite{lopez:data-leakage-llm:arxiv24, deng:benchmark-probing:neurips2023, kaswan:memorisation:icse24, yang:unveiling-memorization:icse24}. This concern arises from the possibility that LLMs might have been exposed to the coding tasks during their training phase, thereby skewing evaluation results. To mitigate the risk of data leakage, it is essential to consider the training cut-off time, which refers to the point in time beyond which data was not included in the model's training set. However, a notable complication is the frequent lack of transparency regarding the exact cut-off times for the training datasets of LLMs, making it challenging to implement this mitigation strategy effectively.


Recently, the ClassEval~\cite{du:classeval:icse2024} benchmark has been introduced, which represents a significant step forward in evaluating class-level code generation capabilities of LLMs. ClassEval was created manually to mitigate potential data leakages from existing code sources. However, it still fails to capture the complexities of APIs and cross-file interactions. Its primary limitation lies in its narrow focus on class skeletons and method-level tasks, without considering how these classes interact within the larger ecosystem of a software project. To bridge this gap, we propose a new benchmark \benchmarkfullname (\benchmark) that uses a handcrafted dataset emphasizing cross-file API usage across various tasks, offering a more comprehensive evaluation of LLMs in complex software development scenarios. Furthermore, ClassEval was specifically developed for Python. In contrast, we develop a multilingual benchmark designed to support both Java and Python, broadening the scope of evaluation to encompass a more diverse set of programming languages.

\header{Benchmark Construction Methodology} 
Our goal was to produce a diverse set of applications, each leveraging ClassEval to fulfill distinct purposes, thereby reflecting the complexity and diversity observed in real-world software development contexts. 
With this goal in mind, we engaged in a brainstorming session involving all the authors to develop potential applications that could leverage the functionalities provided by the ClassEval benchmark. This internal collaborative ideation process ensured that the tasks were reflective of real development challenges. This ideation step resulted in the identification of 14 application development tasks that leverage the underlying functionalities from the ClassEval benchmark. For each application, we begin by manually creating class skeletons and brainstorming the functionalities that this application would provide. 
The benchmark was created by authors with ten years of software development experience each. One author led the development of the Python application, while another focused on creating Java applications. This strategy guaranteed that the benchmark would be applicable across a variety of programming languages. After development, a thorough review process was conducted. This peer review step ensured the quality and functionality of the code. Differences in implementation prompted discussions among the authors to reach a consensus. Key practices involved uniform code formatting and the use of meaningful naming conventions appropriate for each programming language. This process highlighted the collaborative nature of software development. Thus, ground truth solutions were manually developed for each task, serving as a benchmark for evaluating the outputs generated by the LLMs. 

\header{Construction of the Next Token API Dataset}
The construction of this dataset began with the ground truth solutions mutually agreed upon by the two authors. We focused exclusively on API calls made to previously unseen code entities introduced in the APIEval dataset. Common API calls from well-known libraries were deliberately excluded. For instance, we omitted calls related to the \code{Map} functionality, among others. This decision was driven by the aim to simulate a scenario where a developer is working on a proprietary project unfamiliar to the LLM. For each identified API call, we generated a task for the Next Token API completion task. This involved retaining all preceding code from the top of the class up to the point of the API call. 
For an API token completion task shown in Listing~\ref{lst:class-with-cross-file-context}, the ground truth is \code{tp.sentiment\_analysis}, accompanied by a single method parameter \code{review\_text}.

\header{Construction of the Conversational API Dataset}
To simulate queries related to unseen software projects, we initially manually created a query that a developer is likely to write. For instance, as demonstrated in Listing~\ref{lst:class-with-cross-file-context}, queries are unambiguous and provide detailed context, like \code{how to process payment with PaymentProcessor}. The objective is to provide the AI-assisted tool with comprehensive contextual information to assess its understanding during conversational API query completion in a controlled manner. This setup aims to ascertain whether the tool can accurately fulfill query completion tasks based on a clear description of the developer's intention. Furthermore, we added statements related to the API query and then employed \gpt to generate how a developer might phrase this call. This use of \gpt to create additional queries is based on its familiarity with developer interactions during coding tasks, offering a realistic representation of a developer query. For instance, in the case of Listing~\ref{lst:class-with-cross-file-context}, we referenced (line-29) and inquired with \gpt how a developer would instruct \copilot to perform the code completion task. Through these steps, we developed the conversational API completion dataset.

In our example, for the developer query \code{how to process payment with PaymentProcessor?}, the ground truth is taken from the solution as developed in the previous steps. This strategy ensures every query aligns with a specific and relevant API call.

\header{Benchmark Characteristics} 
\benchmark contains a total of 6.9K and 5.5K lines of code for the Python and Java unseen code projects, respectively. \benchmark encompasses a range of functions, such as processing various file formats, performing database operations, executing arithmetic algorithms, handling basic natural language processing tasks, and managing web networking procedures. Based on the functionalities, these dependencies are placed in the corresponding packages that resemble their behaviors. These dependencies consist of 6K and 4.2K lines of code for Python and Java, respectively. Overall, there are 10 Python dependency packages and 8 Java dependency packages. On average, each Python package contains 14.5 files, with each file having 41 lines, and each Java package contains 12 files, with each file having 43 lines.

Based on these functionalities, \benchmark is composed of 14 application classes alongside their associated dependencies written in Python and Java, generated with approximately 120 person-hours of effort. Each application, on average, consists of 76 lines with 1247 tokens and incorporates 7 import statements. Among these, 5 imports are specific to the project while 2 come from established libraries, reflecting the complexity and diversity of real-world software development scenarios. The benchmark comprises \textbf{552} tasks, specifically designed to assess API usage across two categories: \textbf{272} tasks for Next Token API call completions and \textbf{280} tasks for Conversational API completions, offering a varied set of challenges for language models. In comparison, established benchmarks such as HumanEval~\cite{chen:humaneval:arxiv21}, MBPP~\cite{austin:program-synthesis-and-llm:arxiv21}, and ClassEval~\cite{du:classeval:icse2024} contain 164, 974, and 100 tasks, respectively. 

\subsection{Language Models} We employ popular LLMs, which are shown to be effective in the literature with different code-related tasks~\cite{ren:llm-for-code-on-safe-code:arxiv24, kaswan:memorisation:icse24, mcnutt:ai-powered-code-assistants:chi23,ahmed:code-summarization-llm:ase22}.

\textbf{GPT3.5-turbo (GPT-3.5, 2023):} GPT3.5-turbo represents an advanced iteration in the GPT-3 series, developed by OpenAI.

\textbf{GPT-4 (GPT4, 2023):} The GPT-4, also released by OpenAI in 2023, is a more refined version of its predecessor, designed to deliver even greater accuracy and efficiency in code generation and interpretation.


\section{Evaluation}
\label{sec:evaluation}
To assess the effectiveness of \toolname, we address the following research questions: 


\begin{itemize}
\item \textbf{RQ1}: How does \toolname perform on API completion in comparison to prompting LLM with the ad-hoc context?

\item \textbf{RQ2}: How does \toolname's accuracy compare with developer tools?
\end{itemize}

\subsection{Implementation}
\toolname{} is developed in Python, with Tree-sitter~\cite{tree-sitter} facilitating our analysis. For \gptthree and \gpt, we use \code{gpt-3.5-turbo-0125} and \code{gpt-4-0613} , respectively. For all experiments, we set the temperature parameter to 0 to ensure deterministic and well-defined answers from the LLM. Chroma DB, which is an open-source embedding database\footnote{\url{https://www.trychroma.com/}}, serves our need for vector similarity search. For embedding generation, we use \texttt{text-embedding-ada-002} from OpenAI. In the following subsections, we outline the results of the experiments designed to address each of the posed research questions. The \benchmark is designed for both Java and Python as described in Section~\ref{sec:apieval}.

\begin{table*}[h]
    \centering
    \scriptsize
    \caption{Results for the API token completion task on Java and Python}
    \resizebox{\textwidth}{!}{
    \begin{tabular}{l | l | c | c | c}
        \toprule
        \textbf{Programming Language} & \textbf{Tool/Model and Strategy} & \textbf{Call Accuracy (\%)} & \textbf{Argument Matching (\%)} & \textbf{Inference Time (ms)} \\
        \midrule
        \multirow{5}{*}{Java} 
            & \copilot & 58.8 & 53.6 & - \\
            & \gptthree Ad-hoc context & 29.6 & 25.7 & 1365 \\
            & \gpt Ad-hoc context & 47.4 & 44.7 & 1135 \\
            & \toolname with \gptthree & 69.3 & 66.7 & 1220 \\
            & \toolname with \gpt & \textbf{83.7} & \textbf{80.4} & 1342 \\
        \midrule
        \multirow{5}{*}{Python} 
            & \copilot & 57.1 & 52.1 & - \\
            & \gptthree Ad-hoc context & 37.8 & 33.6 & 1099 \\
            & \gpt Ad-hoc context & 43.7 & 36.1 & 1048 \\
            & \toolname with \gptthree & 73.9 & 65.5 & 1069 \\
            & \toolname with \gpt & \textbf{81.5} & \textbf{66.4} & 1176 \\
        \bottomrule
    \end{tabular}
    }
    \label{table:api_token_completion}
\end{table*}

\subsection{Evaluation metrics}
\label{subsec:evaluation-metrics}

\header{Function Call Accuracy (\%)} 
This metric is used to determine the percentage of samples where the inferred output matches with the API function name, disregarding the arguments.

\header{Argument Matching (\%)} 
This metric evaluates when the function call arguments match with the expected output. We take into account cases where the arguments are not expressed exactly as the expected output but will achieve identical outcomes, allowing for different variations in code that accomplish the same task.

\header{Inference time} This is the amount of time in seconds to predict the output given a prompt.

\subsection{Effectiveness of \toolname (RQ1)}
\label{ssec:effectiveprompt}
To study the effectiveness of \toolname, we conducted experiments focusing on both the API next token and conversational query tasks with the \benchmark. The evaluations are performed on Java and Python programming languages with local file context and \toolname. Tables \ref{table:api_token_completion} and \ref{table:results-api_conversation_completion} show the results of these API-completion-related tasks. We discuss the results for each task below.

\header{Next token API completion} In Table~\ref{table:api_token_completion}, we observe the results for API token completion tasks. With the local file context, \gptthree for token completion tasks with Java, the call accuracy is 29.4\%, and the argument matching accuracy is 25.5\%, which is the lowest across all Java token completion tasks. This accuracy increased significantly to 69.3\% for the call accuracy and 66.7\% for argument matching with \toolname. For tasks employing \gpt in Java, call accuracy with the local-file context is observed at 47.1\%, with an argument matching accuracy of 44.4\%, which is significantly higher in comparison with \gptthree with the local-file context. \toolname yielded the best result with \gpt where the call accuracy reached \textbf{83.7\%} and argument matching at \textbf{80.4\%}, which surpassed the best results obtained with local file context by 1.77 times and 1.81 times, respectively. This improvement was observed despite a slight increase in inference time to 1342 milliseconds.

For Python API token completion, the local file context with \gptthree call accuracy and argument matching accuracy are 37.8\% and 33.6\%, respectively, which is the lowest across all Python API token completion tasks. However, it is higher than the accuracy achieved by \gptthree for Java using local file context. The use of \toolname improved call and argument matching accuracy to 73.9\% and 65.5\%, respectively, marking a significant improvement. With \gpt, call accuracy increased to 43.7\%, yet this remains lower than the performance of \toolname with \gptthree. \toolname combined with \gpt achieved the best result for Python, with the highest accuracy of \textbf{81.5\%} in call accuracy and \textbf{66.4\%} in argument matching, the highest recorded in any Python API token completion task, exceeding the corresponding results for the best local file context by 1.86 and 1.84 times, respectively.

\header{Conversational API completion}
The analysis of conversational API completion tasks, presented in Table~\ref{table:results-api_conversation_completion}, compares the performance of various models and strategies in Java and Python programming languages.

For conversational queries with Java, \gptthree with local-file context achieves a call accuracy of 24.4\% and an argument matching accuracy of 11.1\%, with an inference time of 789 milliseconds. This accuracy is the lowest among all Java conversational query completion tasks. \toolname with \gptthree substantially enhances the results to 74.4\% in call accuracy and 63.2\% in argument matching, albeit with a longer inference time of 1011 milliseconds. \gpt with a local file context attains a call accuracy of 38.4\% and an argument matching accuracy of 21.0\%, surpassing the conversational API completion performance with \gptthree. When we leverage \toolname with GPT-4, accuracy increased to \textbf{79.7\%} and \textbf{73.2\%} for call and argument matching, outperforming the best results achieved with local file context by 2.07 times and 3.49 times, respectively. This improvement was observed, notwithstanding a slight increase in inference time to 1145 milliseconds.

Next, we analyze conversation API completion tasks for Python. When employing \gptthree within a local-file context for conversational queries, call accuracy is noted at 35.5\%, and argument matching accuracy at 20.0\%, with an inference time of 718 milliseconds. This performance is notably improved upon introducing \toolname with \gptthree and achieved a call accuracy of \textbf{74.1\%} and argument matching accuracy to \textbf{57.8\%}, surpassing the corresponding results for the best local file context by 2.08 times and 2.89 times, respectively. \gpt with the local-file context in Python achieved the lowest call accuracy of 34.1\% and an argument matching accuracy of 14.8\%, with an inference time of 979 milliseconds. The integration of \toolname with \gpt significantly improves accuracy to 71.9\% for call and 53.3\% for argument matching, respectively.

\subsection{Comparison with State-of-the-Art Developer Tools (RQ2)}
\label{ssec:comparison}
We compare \toolname with \copilot, which is a state-of-the-art developer tool that has shown significant improvements in software developer productivity. 

\header{Setup} For this comparison, we leverage \copilot within GitHub Codespaces in Visual Studio Code\footnote{\url{https://code.visualstudio.com/docs/remote/codespaces}}. We set up a Codespace workspace with \benchmark. This approach facilitates a fair evaluation of the effectiveness of \copilot in scenarios typical of proprietary, company-specific coding projects. In our experimental setup, Codespaces has access to all files present in the project. Our analysis focuses on two main tasks: the API next token completion task, for which we open a file containing an incomplete API call that \copilot is expected to complete by suggesting the next code token. For the conversational API query task, we open a file with a developer query asking to complete an API call, requiring \copilot to assist in completing an API call. Due to \copilot's lack of an inference time metric and satisfactory response latency, we have omitted inference time from our evaluation metrics for \copilot.

\header{Results} As shown in Table~\ref{table:api_token_completion}, for the API token completion task in Java and Python, Copilot achieves a call accuracy of 58.8\% and 57.1\%, respectively. In comparison, the configuration for \toolname achieved 83.7\% in Java and 81.5\% in Python for call accuracy. A similar trend was observed for argument matching, with Copilot securing 53.6\% in Java and 52.1\% in Python, whereas \toolname yielded a higher accuracy of 80.4\% in Java and 66.4\% in Python for argument matching accuracy.

For the conversational API completion tasks, as shown in Table~\ref{table:results-api_conversation_completion}, Copilot's performance in Java and Python demonstrates call accuracies of 50.7\% and 58.6\%, respectively. In comparison, the highest accuracy from \toolname for call accuracy is 79.7\% in Java and 74.1\% in Python. For argument matching accuracy, Copilot achieved 36.2\% in Java and 40.0\% in Python. In contrast, \toolname exhibits higher accuracy with argument matching accuracies of 73.2\% in Java and 57.8\% in Python. Overall, \toolname outperforms \copilot by achieving call exact matches by 1.43 times higher for the token completion and 1.42 times higher for the conversational query completion tasks, respectively.

\begin{table*}[h]
    \centering
    \caption{Results for the conversational API completion task on Java and Python}
    \resizebox{\textwidth}{!}{
    \begin{tabular}{l | l | c | c | c}
        \toprule
        \textbf{Programming Language} & \textbf{Tool/Model and Strategy} & \textbf{Call Accuracy (\%)} & \textbf{Argument Matching (\%)} & \textbf{Inference Time (ms)} \\
        \midrule
        \multirow{5}{*}{Java} 
            & \copilot & 50.7 & 36.2 & - \\
            & \gptthree Ad-hoc context & 24.4 & 11.1 & 789 \\
            & \gpt Ad-hoc context & 38.4 & 21.0 & 1110 \\
            & \toolname with \gptthree & 74.4 & 63.2 & 1011 \\
            & \toolname with \gpt & \textbf{79.7} & \textbf{73.2} & 1145 \\
        \midrule
        \multirow{5}{*}{Python} 
            & \copilot & 58.6 & 40.0 & - \\
            & \gptthree Ad-hoc context & 35.5 & 20.0 & 718 \\
            & \gpt Ad-hoc context & 34.1 & 14.8 & 979 \\            
            &  \toolname with \gptthree & \textbf{74.1} & \textbf{57.8} & 693 \\
            & \toolname with \gpt & 71.9 & 53.3 & 1061 \\
        \bottomrule
    \end{tabular}
    }
    \label{table:results-api_conversation_completion}
\end{table*}

\section{Discussion}

\subsection{Role of Relevant Contextual Information}
\toolname incorporates a lightweight analysis step to gather relevant contextual information, aiding developers in API completion tasks. Our findings indicate that providing the pertinent contextual information is pivotal for the underlying model. For both token completion and conversational query completion tasks, relying solely on local context proves insufficient. In contrast, \toolname, with its context analysis phase, outperforms \copilot by employing project-specific information from cross-file context to improve call accuracy and argument matching. Our approach provides the language model with the necessary contextual guidance, hence enabling the model to generate more precise predictions.

The importance of context also becomes apparent in conversational API completion tasks. LLMs, without all the required context, cannot understand the intent and requirements for completing a given conversational prompt. \toolname instead addresses this gap by implementing a retrieval-based method to align the requirements from a developer query with the underlying repository structure of a project. This contextual query mapping step ensures that \toolname can interpret and respond to API completion query by drawing on relevant project information. Therefore, contextual relevancy in prompts is pivotal in augmenting the effectiveness of LLMs in processing unseen software projects.

\subsection{Qualitative Analysis of API Completion Tasks}
\label{ssec:qualitativeanalysis}
To evaluate the quality and discern the distinct characteristics of outputs generated by various models, we undertook a sampling of the outputs from the models in use, followed by a thorough analysis. The ensuing subsections will present a qualitative examination of these results, complemented by specific examples.


\header{Plausibility of the Generated Outputs}
It was observed that, in the majority of instances, the outputs from \copilot can semantically fulfil the designated tasks. Despite this semantic alignment, the invoked methods were not always congruent with the anticipated API calls. The similarities can be observed in \autoref{lst:semantic-similarity}. 

\begin{lstlisting}[language=Python, caption={Semantic Similarity of Failed Completions},label={lst:semantic-similarity}]
is_course_active(course_id)    // expected
is_course_available(course_id) // generated

validate_email_address(user_details.email) // expected
is_valid_email(user_details['email'])      // generated

display_items() // expected
get_all_items() // generated
\end{lstlisting}

In contrast, \toolname successfully generated the correct API calls for these instances. During the context analysis phase, we collect all the available methods and then provide them during the prompting phase, which equips the model with an understanding of potential API calls.

\header{Lack of Context for Completion} In instances where the intended action within the code is ambiguous, given the overall context, multiple potential actions may be identified. Under these circumstances, both \toolname and alternative methods face challenges in accurately generating the correct API call during next-token prediction tasks. Specifically, \toolname has demonstrated the capacity to produce valid API calls by leveraging the module utilized for call generation. However, these calls occasionally do not align with the expected function at the designated point in the code. This issue is notably less pronounced in the context of conversational API completion tasks, where the clarity of the tasks tends to reduce the occurrence of mismatches. \autoref{lst:amb-ctx} shows a few cases of unintended valid API calls because of the context ambiguity.

\begin{lstlisting}[language=Python, caption={Ambiguous Completion Context},label={lst:amb-ctx}]
chat.add_user("VendingBot") // expected
chat.send_message(sender, receiver, message) // generated

player.add_song(recommended_song) // expected
set_current_song(recommended_song['title']) // generated
\end{lstlisting}

However, the occurrence of such cases is not that high, leading to the observation that overall accuracy for conversational completions is lower than next token completion.

\subsection{Development of Context-aware Coding AI-Assistance Tools}
\label{ssec:development_context_aware_tools}
Our results demonstrate the need to embed contextual understanding within AI-assisted tools designed to boost developer productivity. Although existing tools are capable of aiding developers with well-known libraries, their effectiveness is hindered in a company-specific code base, largely due to the lack of project-specific domain knowledge. This lack of context leads to the generation of plausible yet incorrect code suggestions, which could potentially impair the adoption of these tools. Our findings highlight that while \copilot is a powerful tool, it falls short in aligning developer intentions with specific project contexts. As the adoption of conversational AI becomes more prevalent, making sense of a query from developers becomes pivotal.

To mitigate this, \toolname builds a project-specific database and makes it amenable to an embedding-based search. This helps to map the developer query to the underlying project structure for an unseen query without incorporating additional latency. On average, a retrieval-based query takes 200 milliseconds. As conversational AI becomes more prevalent, this notion of retrieval-based mapping of repository code-structure mapping employed in \toolname through a query understanding phase can play a role.
\section{Threats To Validity}

\header{Benchmark size and generalizability} 
A potential threat is the risk of data leakage between our benchmark and the training data of the models under investigation. To mitigate this risk, we have taken a deliberate approach by manually constructing \benchmark. This manual construction process ensures that the tasks included in the benchmark are not directly drawn from any pre-existing datasets known to have been used in the training of the evaluated LLMs. The current size of \benchmark, may not fully ensure the generalizability of our findings across a wider array of software development contexts and programming languages. To mitigate this threat, we have used the common coding patterns and practices used by developers to reuse code across different modules. In the future, we aim to expand this initial benchmark with more varied and complex project contexts.

\header{Benchmark construction bias}
One potential threat to the validity of our approach lies in the manual construction of the evaluation benchmark. To mitigate this threat, \benchmark builds upon a previously established benchmark ClassEval~\cite{du:classeval:icse2024} that closely mimics the complexity and intricacies of real-world object-oriented programming scenarios.
Additionally, the authors adhered to best practices in developing applications so that they closely resemble realistic scenarios.

\header{Choice of the conversational query}
The formulation of conversational queries presented to AI-assisted tools such as \copilot for completing API tasks can play a role in the outcomes of our study. To address this, we have provided detailed guidance within these queries, aiming to clarify the intention of the developer for the API completion task. A potential limitation stems from the diversity in how developers may interact with \copilot with their queries in real-world scenarios. The rationale behind providing detailed conversational queries is to establish a controlled environment in which the explicit intention of the developer is communicated to \copilot. Our objective was if unambiguous developer intention is provided with detailed guidance, how well \copilot would work. This approach allows us to isolate and evaluate the effectiveness of \copilot in understanding and executing API completion tasks when given clear and detailed instructions.

\header{Scope of programming language} Although \benchmark is designed to be multilingual, focusing on Java and Python, the generalizability of our findings to other programming languages or paradigms, such as functional programming languages, is not guaranteed. Different languages may pose unique challenges for API completion that our benchmark has not addressed. However, there are existing benchmarks that focus on a single programming language~\cite{chen:humaneval:arxiv21, austin:program-synthesis-and-llm:arxiv21, du:classeval:icse2024}. In this study, we focus on Java and Python due to their popularity and widespread use in the community~\cite{tiobe-language-popularity-index}.

\header{Large language models on source code} In this study, we evaluated the effectiveness of \toolname with 2 different LLMs. There are other language models on source code namely \textsc{CodeLlama\xspace}~\cite{codellama:arxiv23}, \incoder~\cite{fried:incoder:arxiv22}, \codetfive~\cite{wang:codet5:acl21}, among others~\cite{nijkamp:codegen:arxiv22, feng:codebert:emnlp20, guo:unixcoder:acl22, guo:graphcodebert:iclr21, lu:codegpt:arxiv21, jain:contrastive-learning:acl21, phan:cotext:nlp4prog21, plbart:ahmad:acl21, xu:polycoder:maps22}. While incorporating a wider array of LLMs into our empirical evaluation could be useful, the primary focus of this paper is not to conduct exhaustive comparisons across all available models. In this work, we investigate how to leverage project-specific context for API completion tasks, and we have included LLMs that are widely studied in the academic community~\cite{xie:chatunitest:arxiv23, yuan:evaluate-chatgpt:arxiv2023, bang2:hallucination-multitask:23, tian:chatgpt-programming-assistance:arxiv2023}.

\section{Related Work}

\header{Reliability Detection for LLMs} 
LLMs often generate incorrect or non-factual statements, with factors like data quality~\cite{lee-etal-2022-deduplicating}, source-target divergence~\cite{dhingra:evaluating-table-to-text-generation:acl2019}, and randomness during inference~\cite{dziri-etal-2021-neural} contributing to this issue~\cite{bang2:hallucination-multitask:23, dai:plausible-may-not-be-faithful:23}. 
Some studies detect hallucinations using logit outputs or internal states, but these methods have limitations, particularly in black-box scenarios such as ChatGPT~\cite{manakul-etal-2023-selfcheckgpt}. 
Unlike existing approaches that either modify LLMs or analyze output inconsistencies, our \toolname tool enhances reliability by incorporating relevant context without needing LLM internal states.

\header{Applications of language models on code generation} 
Recently, LLMs have been applied to different stages of software development, such as code snippet generation~\cite{openai:gpt4:arXiv2023}, test code generation~\cite{xie:chatunitest:arxiv23}, and code repair~\cite{xia:chatrepair:arxiv2023}.
There are also approaches to include repository-level information~\cite{shrivastava:repo-level-code:icml23, ding:cocomic:arxiv22} 
to retrieve relevant contexts for statement completion. Both approaches modify the LLM's operation, either through loss function calculation~\cite{shrivastava:repo-level-code:icml23} or fine-tuning~\cite{ding:cocomic:arxiv22}. In contrast, \toolname introduces an LLM-agnostic approach without modifying underlying LLM by providing adequate contextual information for the API completion task. A recent work~\cite{phan:repohyper:arxiv24} addresses repository-level code completion for Python by training a Link Predictor model. 
In contrast, our approach can be directly applied as a black-box solution to the underlying LLM without the prerequisite of training and maintaining a model. 
Additionally, we support conversational queries within a repository-specific context in a multilingual environment.

\header{Library focused API recommendation and code generation}
There are API recommendation approaches that employ retrieval-based~\cite{buse:retrieval-based-synthesizing-api-usage:icse12, gvero:retrieval-based:synthesizing-java-expressions:sigplan15, huang:retrieval-based-api-method-recommendation:ase18, liu:retrieval-based-generating-query-specific:fse19, mcmillan:retrieval-based-finding-relevant-functions:icse11, raghothaman:retrieval-based-synthesizing-what-i-mean:icse11, rahman:retrieval-based-nlp2api:icsme18, rahman:retrieval-based-rack:saner16} and learning-based~\cite{gu:deep-api-learning:fse16, hadi:pretrained-models-for-api-learning:icpc22, martin:deep-api-learning:icpc22} approaches. As an example, there are techniques~\cite{huang:retrieval-based-api-method-recommendation:ase18, wei:clear:icse22} to recommend API based on similar Stack Overflow posts. Learning-based techniques~\cite{gu:deep-api-learning:fse16, hadi:pretrained-models-for-api-learning:icpc22, martin:deep-api-learning:icpc22} employ pre-trained models for API recommendation. However, these approaches rely on historical code usage and forum data, demonstrating limitations when encountering software projects not previously seen in the dataset. Works in code generation have lately employed retrieval-based methods to use APIs from a given library~\cite{liu:retrieval-based:ase23, shuyan:retrieval-based-docprompting:iclr23, zan:retrieval-based:emnlp2022} using LLMs. However, these approaches are based on the assumption that libraries have elaborate API documentation available, a condition that may not be universally met, potentially limiting their applicability. A recent effort~\cite{liu:retrieval-based:ase23} leverages previous similar samples from the training set, which makes it inapplicable to unseen software projects that are not present in the training data. 

Pei et al.~\cite{pei:better-context-function-call-argument-completion:aaai23} focuses on function call argument completion by leveraging non-local information with the help of Language Server Protocol~\footnote{https://microsoft.github.io/language-server-protocol} specific to Python. In contrast, our context analysis is based on language-agnostic AST~\cite{tree-sitter} and can support conversation API completion tasks, which is beyond the scope of function call argument completion.

\header{Code benchmarks}
In section~\ref{sec:apieval}, we presented the most recent works on LLMs and existing code generation benchmarks. 
There are also task-specific benchmarks for LLM evaluation. For example, Spider~\cite{yu:spider-text-to-sql:acl18} is used to assess database query capabilities, while DS-1000~\cite{lai:ds-1000-data-science-code-benchmark:pmlr23} evaluates performance in data science scenarios. A recent work~\cite{yu:codereval:icse24} introduces a benchmark that focuses on function-level code generation from natural language descriptions curated from GitHub projects. This approach, while valuable for assessing code generation capabilities, may be susceptible to data leakage, as models might have been previously exposed to similar or identical code during their training on public code repositories. 
In contrast, \benchmark includes a more realistic scenario for API completion task, where developers are tasked with completing code in the context of an unseen software project. 
\section{Conclusion}
This paper introduces \toolname, an approach to incorporate intra-repository context for improving API completion tasks. Through the analysis of import statements and retrieval-based mapping, \toolname aligns developer intentions with the most contextually relevant API suggestions. We introduce \benchmark, a multilingual benchmark, specifically tailored for evaluating cross-file API invocation tasks with LLMs across Java and Python projects. 
Our evaluation of \toolname shows that it can achieve call exact matches with 82.6\% and 76.9\% accuracy for API next token and conversational query tasks, respectively. 
Notably, \toolname outperforms the leading developer tool, \copilot, in terms of call accuracy and argument matching. This highlights the critical role of contextual information in addressing API completion tasks and overcoming the obstacles posed by a lack of domain-specific knowledge within proprietary codebases. 
As part of our future work, we plan to extend this notion of lightweight context analysis to repository-specific test code generation. We also plan to incorporate our approach as an IDE plugin to assist developers with their software engineering tasks.

\section{Data Availability}
Paper is currently under review. Code artifacts will be made available soon.

\bibliographystyle{IEEEtran}
\interlinepenalty=10000
\bibliography{learning-on-source-code-processing-tasks}

\end{document}